\documentclass[aps,showpacs,pre,superscriptaddress,eqsecnum,twocolumn]{revtex4}
\usepackage{hyperref}
\usepackage{amsmath}
\usepackage{amssymb}
\usepackage{epsfig}
\usepackage{natbib}
\newcommand*{\be}{\begin{equation}}
\newcommand*{\ee}{\end{equation}}
\begin{document}
%to be published in Phys. Rev. E
\bibliographystyle{revtex}
\title[Ablowitz-Ladik soliton radiation]{Perturbation-induced radiation
by the Ablowitz-Ladik soliton}
%\date{\today}
\author{E.V. Doktorov}
\email{doktorov@dragon.bas-net.by} \affiliation{B.I. Stepanov
Institute of Physics, 68 F. Skaryna Ave., 220072 Minsk, Belarus}
\author{N.P. Matsuka}
\email{matsuka@im.bas-net.by}
\affiliation{Institute of
Mathematics, 11 Surganov Str., 220072 Minsk, Belarus}
\author{V.M. Rothos}
\email{v.m.rothos@qmul.ac.uk} \affiliation{School of Mathematical
Sciences, Queen Mary College, University of London, Mile End Road,
London E1 4NS, U.K.}

\begin{abstract}
An efficient formalism is elaborated to analytically describe
dynamics of the Ablowitz-Ladik soliton in the presence of
perturbations. This formalism is based on using the
Riemann-Hilbert problem and provides the means of calculating
evolution of the discrete soliton parameters, as well as shape
distortion and perturbation-induced radiation effects. As an
example, soliton characteristics are calculated for linear damping
and quintic perturbations.
\end{abstract}

 \maketitle

\section{Introduction}

Dynamics of discrete solitons (intrinsic localized modes) in
nonlinear lattices has become a topic of intense research
summarized in a number of excellent reviews \cite{Rev-1}.
Propagation properties of waves arising as a result of the
interplay of nonlinearity with lattice discreteness can be quite
distinct from those inherent in continuous nonlinear systems and
hold much promise for applications in various physical, biological
and technological problems. Examples are energy localization and
transfer in systems of nonlinear oscillators \cite{Eil},
propagation of self-trapped beams in arrays of coupled nonlinear
optical waveguides \cite{Ch-J,Exp-1}, nonlinear charge and
excitation transport in biological macromolecules
\cite{Dav,Scott}, local denaturation of DNA double helix
\cite{Pey}, dynamics of localized excitations in arrays of coupled
Josephson junctions \cite{Ust}, propagation of optical spatial
solitons in nonlinear photonic crystals \cite{Ming} and in
diffraction-managed waveguide systems \cite{Exp-2}, creating
discrete solitons in Bose-Einstein condensate \cite{Trom}.
Recently it was proposed \cite{Eugen} to use discrete solitons in
two-dimensional networks of nonlinear waveguides to realize
functional operations such as blocking, routing, logic functions,
and time gating.

Most of the above phenomena are modelled by the discrete nonlinear
Schr\"odinger (DNLS) equation or, in a more general setting, by
the discrete self-trapping equation \cite{Eil}. Recent
developments in the study of the DNLS equation are reviewed in
Refs~\cite{Rev-2,Rev-3}. However, the standard DNLS equation is
nonintegrable \cite{Sharf,Herbst} and does not exhibit exact
soliton solutions, though it can be derived as a discretization of
the integrable continuous NLS equation. Hence, numerical methods
are generally used  to investigate nonlinear lattice dynamics on
the basis of the DNLS equation.

On the other hand,  there exists the integrable discretization of
the NLS equation - the Ablowitz-Ladik (AL) equation \cite{AL}
which has exact soliton solutions and admits the complete
description in the framework of the inverse spectral method.
Moreover, Konotop \textit {et al}. \cite{Kon} and Cai \textit {et
al}. \cite{Cai-1} proved integrability of the inhomogeneous AL
system in an external electric field of a particular form. Being
unique from the mathematical point of view, the AL equation is
less applicable in physics than the DNLS equation. Salerno
\cite{Sal} introduced an equation that interpolates between the
DNLS and AL equations and permits studying (as a rule,
numerically) the role of integrable and non-integrable
contributions to lattice properties \cite{Cai-2}. The AL-DNLS
system with an impurity was investigated by Hennig \textit {et
al}. \cite{Hen}

A different point of view on the interrelation between the AL and
DNLS equations was posed in Refs.~\cite{V-G,Kiv-1,Acev}. In a
definite region of parameters the DNLS equation can be treated as
a perturbed version of the AL equation. When a perturbation is
small, the discrete soliton perturbation theory can be
successfully applied to analytically describe localized
excitations in a system governed by the DNLS equation. Such an
approach was developed in Refs.~\cite{V-G,Kiv-1,Acev} in the
framework of the adiabatic approximation, when a
perturbation-induced radiation is ignored and a perturbation
manifests itself as a slow evolution of initially constant AL
soliton parameters. The evolution equations for the parameters
were derived by Vakhnenko and Gaididei ~\cite{V-G}. Stability
aspects of Hamiltonian perturbations for the AL equation were
discussed by Kapitula and Kevrekidis~\cite{Kap}. Recently the
perturbative method to study the AL soliton dynamics was used in
Ref.~\cite{Kundu} in relation to energy transport in
$\alpha$-helical proteins and in Ref.~\cite{Gar} for the soliton
in a random medium. Besides, Abdullaev \textit {et al.} \cite{Abd}
proved the existence of discrete autosolitons in the AL model with
linear and quintic damping, cubic amplification and complex
filtering (the discrete complex Ginzburg-Landau model). Exact
solutions of this model for certain relations between parameters
are given in Ref.~\cite{Mar}.

It is well known that a perturbation of the soliton is also
accompanied  by radiation of small-amplitude dispersive linear
waves (or shape distortion)~\cite{K-M}, and a complete description
of the perturbed soliton dynamics necessitates accounting for both
the soliton parameter evolution and the radiation effects.
Therefore, the main goal of the present paper is to develop a
corresponding (relatively simple) formalism and to extend, as far
as possible, the applicability of analytical methods in studying
near-integrable nonlinear discrete systems. It should be noted in
this connection that Konotop \textit {et al.} \cite{Konot} derived
by means of the Gel'fand-Levitan-like summation equations the
evolution equation for the reflection coefficient in the case of
the inhomogeneous AL model but without using it for specific
calculations. An estimation of radiative corrections to the AL
soliton subjected to the stochastic perturbation was outlined in
the important paper by Garnier~\cite{Gar} on the basis of
conserved quantities.

Our approach utilizes the Riemann-Hilbert (RH) problem~\cite{Zak}.
The  application of the RH problem to perturbed nonlinear
equations was initiated by Kivshar~\cite{Kiv-2} on an example of
the Landau-Lifshitz equation. A purely algebraical calculation of
higher-order corrections to the perturbed NLS soliton and of
radiation effects for a soliton in a doped fiber was performed on
the basis of the RH problem in Ref.~\cite{D}. Such an approach has
been proved to be efficient for a wide class of continuous
perturbed nonlinear equations, including multicomponent
ones~\cite{My}.

This paper gives a self-contained exposition of the AL soliton
perturbation theory. In Sec. \ref{sec:2} we fix preliminary facts
concerning the AL spectral problem which are used in Sec.
\ref{sec:3} to formulate the RH problem. In Sec. \ref{sec:4} we
describe a procedure to solve the RH problem with zeros and obtain
immediately the AL soliton solution in Sec. \ref{sec:5}. We stress
that calculations within the RH problem do not use discrete
analogs of the Gel'fand-Levitan integral equations. Section
\ref{sec:6} is devoted to derivation of the evolution equations
for the RH problem data associated with the AL soliton parameters.
These equations exactly account for the perturbation and  serve in
the subsequent Sections as the generating equations for the
perturbative expansion. Sec. \ref{sec:7} contains brief exposition
of the adiabatic approximation, whereas Sec.~\ref{sec:8}
represents the main result of the paper - derivation of  formulas
for calculating radiative corrections from the continuous part of
the RH problem data. In Sec.~\ref{sec:9} we illustrate the
formalism by the examples  of linear damping and quintic
perturbations.  Appendices contain some technical details of the
applications of the RH problem.

\section{The Ablowitz-Ladik spectral problem}\label{sec:2}
\subsection{Jost solutions and asymptotics}

Integrable discretized NLS equation (AL equation) \be\label{AL}
iu_{nt}+u_{n+1}+u_{n-1}-2u_n+|u_n|^2(u_{n+1}+u_{n-1})=0 \ee for a
scalar complex function $u$ which depends on discrete variable
$n$, $-\infty<n<\infty$, and time $t$ admits the Lax
representation with the AL spectral problem \cite{AL}
\be\label{AL-spec} J(n+1)=(E+Q_n)J(n)E^{-1}, \ee
\[
Q_n=\left(\begin{array}{cc} 0 & u_n \\
-u_n^* & 0 \end{array}\right), \qquad E=\left(\begin{array}{cc}
z&0\\ 0&z^{-1}\end{array}\right),
\]
and the evolutionary equation (subscript $t$ means time
derivative)
\begin{eqnarray}\label{AL-temp}
J_t(n)&=&V(n)J(n)-J(n)\Omega(z),\\
V(n)&=&i\left(\begin{array}{cc} u^*_{n-1}u_n & zu_n-z^{-1}u_{n+1} \\
z^{-1}u_n^*-zu^*_{n-1} &
-u_{n-1}u^*_n\end{array}\right)+\Omega,\nonumber \\
\Omega(z)&=&\frac{i}{2}(z-z^{-1})^2\sigma_3.\nonumber
\end{eqnarray}
It means that Eq.~(\ref{AL}) arises as a compatibility condition
for Eqs.~(\ref{AL-spec}) and~(\ref{AL-temp}). Here $z$ is a
constant spectral parameter and the star stands for the complex
conjugation. The spectral problem in the form~(\ref{AL-spec})
differs slightly from the usual one~\cite{AL} and permits
introducing matrix Jost solutions $J_\pm(n)$  of
Eq.~(\ref{AL-spec}) with the unit asymptotics,
$J_\pm(n)\to\openone$ as $n\to\pm\infty$. $J_\pm(n)$ solve
Eq.~(\ref{AL-temp}) as well. The scattering matrix $S(z)$ defined
by \be\label{scatt} J_-(n)=J_+(n)E^nS(z)E^{-n} \ee has the
structure
\[
S(z)=\left(\begin{array}{rr}a_+ & -b_- \\ b_+ & a_-
\end{array}\right).
\]
The Jost solutions obey the conjugation condition \be\label{conj}
J_\pm^\dagger(n,\bar z)=v_\pm(n)J_\pm^{-1}(n,z), \ee where $\bar
z=1/z^*$, '$\dagger$' means the Hermitean conjugation and
\[
v_+(n)=\prod_{l=n}^\infty\rho_l^{-1}, \qquad
v_-(n)=\prod_{l=-\infty}^{n-1}\rho_l, \qquad \rho_l=1+|u_l|^2.
\]
We also obtain  that $\mathrm{det}J_\pm(n,z)=v_\pm(n)$,
$\mathrm{det}S=v$, where $v=\prod_{n=-\infty}^\infty\rho_n$ and
evidently $v_+(n)v=v_-(n)$. From Eqs.~(\ref{scatt})
and~(\ref{conj}) we obtain $S^\dagger(\bar z)=vS^{-1}(z)$ which
gives $ a_+^*(\bar z)=a_-(z)$, $b_+^*(\bar z)=b_-(z)$.

The AL spectral problem~(\ref{AL-spec}) obeys the important
symmetry ('${\mathcal{P}}$-parity'): if $J(n,z)$ is a solution,
then \be\label{parity}
\mathcal{P}J(n,z)\equiv\sigma_3J(n,-z)\sigma_3\ee is a solution,
too. It follows from Eq.~(\ref{parity}) that diagonal elements of
$J(n,z)$ are even functions of $z$, while off-diagonal entries are
odd functions. The same symmetry is valid for the Jost solutions
and the matrix $S$, the latter means $a_\pm(z)=a_\pm(-z)$,
$b_\pm(z)=-b_\pm(-z)$.

Now we consider asymptotic behavior of the solution $J(n,z)$ for
$z\to\infty$. Let
\[
J(n,z)=J^{(0)}(n)+z^{-1}J^{(1)}(n)+O(z^{-2}), \qquad z\to\infty.
\]
Inserting this expansion into Eq.~(\ref{AL-spec}) gives
\be\label{as-1} J^{(0)}(n+1)=\left(\begin{array}{ll}1& 0\\ 0 &
\rho_n\end{array}\right)J^{(0)}(n),\ee
while the potential $u_n$
is retrieved as \be\label{recon} u_n=-J^{(1)}_{12}/J^{(0)}_{22}.
\ee Note that the asymptotics~(\ref{as-1}) is consistent with the
$\mathcal{P}$-parity property~(\ref{parity}). Similar results hold
for $z\to 0$, when $J(n,z)=J_{(0)}(n)+zJ_{(1)}(n)+O(z^2)$:
\[
J_{(0)}(n+1)=\left(\begin{array}{ll}\rho_n & 0\\ 0 &
1\end{array}\right)J_{(0)}(n).
\]

\subsection{Analyticity}
Let $\mathcal{C}_\pm$ are domains in the complex $z$-plane lying
outside $(+)$ and inside $(-)$ the unit circle $|z|=1$. It follows
from the spectral problem~(\ref{AL-spec}) that the first column
$J_-^{[1]}(n,z)$ of the Jost function $J_-$ and the second one
$J_+^{[2]}(n,z)$ of $J_+$ are analytical in $\mathcal{C}_+$ (and
continuous for $z\to 1$). Hence, the matrix function
\[
\Psi_+(n,z)=\left(J_-^{[1]},J_+^{[2]}\right)(n,z)
\]
is a solution of the spectral problem~(\ref{AL-spec}) and
analytical as a whole in $\mathcal{C}_+$. We obtain from the
conjugation formula Eq.~(\ref{conj}) that the  rows
$(J_-)_{[1]}^{-1}$ and $(J_+)_{[2]}^{-1}$ are analytical in
$\mathcal{C}_-$.  As a result,  the matrix function
\[
\Psi_-^{-1}(n,z)=\left(\begin{array}{l}(J_-)_{[1]}^{-1}\\(J_+)_{[2]}^{-1}\end
{array}\right)(n,z)
\]
 is analytical as a whole in $\mathcal{C}_-$ and solves the
adjoint spectral problem.

Analytical solutions can be expressed in terms of the Jost
functions. Indeed,
\begin{eqnarray}
\Psi_+&=&\left(J_-^{[1]},J_+^{[2]}\right)=\left(a_+J_+^{[1]}+z^{-2n}b_+J_+^{[2]},J_+^{[2]}
\right)\nonumber\\
&=&J_+E^nS_+E^{-n}, \qquad S_+(z)=\left(\begin{array}{ll}a_+ & 0\\
b_+ & 1\end{array}\right),
\label{s+}\end{eqnarray} as well as
\begin{eqnarray*}
\Psi_+&=&J_-E^nS_-E^{-n}, \quad S_-=\left(\begin{array}{cc} 1 &
b_-/v \\ 0 & a_+/v \end{array}\right), \\ S_+&=&SS_-.
\end{eqnarray*}

It follows from these formulas that \be\label{det+}
\det\Psi_+(n,z)=v_+(n)a_+(z). \ee In the same way we obtain
\be\label{det-}
\Psi_-^{-1}=E^nT_+E^{-n}J_+^{-1}=E^nT_-E^{-n}J_-^{-1}, \ee
\[
T_+=\left(\begin{array}{cc} a_-/v & b_-/v \\ 0 & 1
\end{array}\right), \qquad T_-=\left(\begin{array}{ll} 1 & 0\\ b_+ &
a_-\end{array}\right),
\]
\[
\det\Psi_-^{-1}=v_-^{-1}(n)a_-(z), \qquad T_+S=T_-.
\]
Asymptotic behavior of analytical solutions is derived directly
from that of the Jost functions and Eqs.~(\ref{s+}) and~
(\ref{det-}): \begin{eqnarray}\label{as-2}
\Psi_+(n,z)&\longrightarrow&\left(\begin{array}{cc}1 & 0\\
0 & v_+(n)\end{array}\right), \quad z\to\infty, \\
\Psi_-^{-1}(n,z)&\longrightarrow&
\left(\begin{array}{cc}v_-^{-1}(n) & 0\\ 0 & 1\end{array}\right),
\quad z\to0. \nonumber\end{eqnarray} Hence, $\det\Psi_+\rightarrow
v_+(n)$ as $z\to\infty$ which gives from Eq.~(\ref{det+})
$a_+(z)\rightarrow 1$ as $z\to\infty$. Similarly, $a_-(z)\to1$ as
$z\to0$. The conjugation formula for the analytical solutions
follows from Eq.~(\ref{conj}) : \be\label{B}
\Psi_+^\dagger(n,z)=B(n)\Psi_-^{-1}(n,\bar z), \quad
 B(n)=\left(\begin{array}{cc} v_-(n) & 0\\ 0 &
v_+(n)\end{array}\right). \ee

\section{Matrix Riemann-Hilbert problem}\label{sec:3}

Having matrix functions $\Psi_+$ and $\Psi_-^{-1}$ which are
analytical in two complementary domains $\mathcal{C}_+$ and
$\mathcal{C}_-$ of the $z$-plane and continuous on the contour
$|z|=1$, we can pose the matrix Riemann-Hilbert (RH) problem
\be\label{RH-1} \Psi_-^{-1}(n,z)\Psi_+(n,z)=E^nG(z)E^{-n}, \qquad
|z|=1 \ee   as a problem of analytical factorization of the
 matrix function $G(z)$ defined on the unit circle
$|z|=1$. It follows from Eqs.~(\ref{s+}) and~(\ref{det-}) that the
matrix $G$ has the form \be\label{RH-2}
G=T_+S_+=T_-S_-=\left(\begin{array}{cc} 1 & b_-/v
\\ b_+ & 1\end{array}\right).
\ee The normalization of the RH problem is given by
Eq.~(\ref{as-2}).

The RH problem~(\ref{RH-1}) has a non-canonical normalization
depending on the potential $u_n$. It has been proved \cite{Ger}
that it is possible to reformulate the AL spectral
problem~(\ref{AL-spec}) so as to arrive at the RH problem with the
canonical normalization and to give a Hamiltonian formulation with
the canonical Poisson brackets. However, the above canonicity is
achieved at the cost of \textit {nonlinear} dependence of the
spectral problem on the potential. Being useful for treating
non-perturbative AL equation and its integrable generalizations,
such an approach seems to be of less value for the case of
perturbations.

In general, the matrices $\Psi_+$ and $\Psi_-^{-1}$  have zeros in
some points $z_j$ and $\bar z_k$ in their regions of analyticity,
i.e., $\det\Psi_+(z_j)=0$, $z_j\in\mathcal{C}_+$,
$j=1,2,\ldots,N_+$, and $\det\Psi_-^{-1}(\bar z_k)=0$, $\bar
z_k\in\mathcal{C}_-$, $k=1,2,\ldots,N_-$. We suppose that all
zeros are simple and of finite number with $N_+=N_-\equiv N$ (in
other words, we have zero-index RH problem). In virtue of the
$\mathcal{P}$-parity, zeros appear in pairs as $\pm z_j$ and
$\pm\bar z_k$. Taking into account Eqs.~(\ref{det+})
and~(\ref{det-}), we conclude that zeros of $\Psi_+$ and
$\Psi_-^{-1}$ are given by zeros of the scattering matrix
elements: $a_+(\pm z_j)=0$ and $a_-(\pm\bar z_k)=0$.

\section{Regularization of the Riemann-Hilbert
problem}\label{sec:4}

We will solve the RH problem~(\ref{RH-1}) with zeros by means of
its regularization. This procedure consists in extracting from
$\Psi_\pm$ rational factors which are responsible for the
existence of zeros. Indeed, near the point $z_j$ we have
$\det\Psi_+(z)\sim(z-z_j)$. Let us introduce a rational matrix
function $\Xi_j^{-1}=\openone+(z_j-\bar z_j)(z-z_j)^{-1}P_j$,
where $P_j$ is a rank $1$ projector, $P_j^2=P_j$. Because
$P_j=\mathrm{diag}(1,0)$ in an appropriate basis, we obtain
$\det\Xi_j^{-1}=(z-\bar z_j)(z-z_j)^{-1}$. Therefore, the product
$\Psi_+(z)\Xi_j^{-1}(z)$ is regular in the point $z_j$ (its
determinant is nonzero in this point). Regularization of the zero
$-z_j$ is given by a rational function
$\Xi_{-j}^{-1}=\openone-(z_j-\bar z_j)(z+z_j)^{-1}P_{-j}$. As a
result, the matrix function
$\psi_+(n,z)=\Psi_+(n,z)\Xi_j^{-1}\Xi_{-j}^{-1}$ is regular in the
points $\pm z_j$. In the same manner we regularize the matrix
$\Psi_-^{-1}(n,z)$ with zeros in $\pm\bar z_k$. Namely, the matrix
\[
\psi_-^{-1}(n,z)=\Xi_{-k}\Xi_k\Psi_-^{-1}(n,z)
\]
is regular in the points $\pm\bar z_k$ and
\[
\Xi_k=\openone-\frac{z_k-\bar z_k}{z-\bar z_k}P_k, \qquad \Xi_{-k}
=\openone+\frac{z_k-\bar z_k}{z+\bar z_k}P_{-k}.
\]
Regularizing all $4N$ zeros of the RH problem~(\ref{RH-1}), we
represent the functions $\Psi_\pm$ as a product \be\label{RH-3}
\Psi_\pm=\psi_\pm\Gamma \ee of the rational matrix function
\be\label{Gamma}
\Gamma(n,z)=\Xi_{-N}\Xi_N\Xi_{-(N-1)}\Xi_{N-1}\ldots\Xi_{-1}\Xi_1
\ee and the holomorphic matrix functions $\psi_\pm(n,z)$ which
solve the regular RH problem (i.e., without zeros):
\be\label{RH-4}
\psi_-^{-1}(n,z)\psi_+(n,z)=\Gamma(n,z)E^nG(z)E^{-n}\Gamma^{-1}(n,z)
. \ee

The appearance of a simple zero $z_j$ of the matrix $\Psi_+$ means
that there exists an eigenvector $|j\rangle$ with zero eigenvalue,
\be\label{vec-1} \Psi_+(n,z_j)|j\rangle=0. \ee Taking the
Hermitean conjugation of this equality with account of the
conjugation property~(\ref{B}), we obtain $\langle
j|B\Psi_-^{-1}(n,\bar z_j)=0$ with $\langle j|=|j\rangle^\dagger$.
Therefore, the projector $P_j$ can be naturally defined as
\be\label{proj} P_j=\frac{|j\rangle\langle j|B}{\langle
j|B|j\rangle}. \ee For the zero $-z_j$ we have
$\Psi_+(n,-z_j)|-j\rangle=0$. In virtue of the
$\mathcal{P}$-parity, both vectors $\vert j\rangle$ and
$\vert-j\rangle$ are interrelated, $|-j\rangle=\sigma_3|j\rangle$,
and therefore $P_{-j}=\sigma_3P_j\sigma_3$.

For practical purposes, it is more convenient to decompose the
products~(\ref{Gamma}) into simple fractions. Following Refs.~
\cite{Kaw,DR}, we obtain \be\label{Gamma-1}
\Gamma(n,z)=\openone-\sum_{j,k=1}^{2N}\frac{1}{z-\bar
z_k}|y_j\rangle(D^{-1})_{jk}\langle y_k|B, \ee
\[
\Gamma^{-1}(n,z)=\openone +\sum_{j,k=1}^{2N}\frac{1}{z-
z_j}|y_j\rangle(D^{-1})_{jk}\langle y_k|B \] with new vectors
$|y_j\rangle$, where zeros are enumerated as
$z_1,-z_1,z_2,-z_2,\ldots,z_N,-z_N$, (and similarly for $\pm\bar
z_k$), whereas matrix elements $D_{kj}$ are given by \be\label{D}
D_{kj}=\langle y_k|\frac{B}{z_j-\bar z_k}|y_j\rangle. \ee It is
seen from Eq.~(\ref{Gamma-1}) that the asymptotic expansion for
$\Gamma(n,z)$ has the form
\[
\Gamma(n,z)=\openone+z^{-1}\Gamma^{(1)}(n)+O(z^{-2}).
\]
Because
$\Psi_+=\Psi_+^{(0)}+z^{-1}\Psi_+^{(1)}+O(z^{-2})=\psi_+(\openone+z^{-1}\Gamma^{(1)}+O(z^{-2})$,
we can choose a $z$-independent function  $\psi_+$ as a solution
of the regular RH problem~(\ref{RH-4}): \be\label{RH-reg}
\psi_+(n)=\Psi_+^{(0)}(n)=\left(\begin{array}{cc}1 & 0\\ 0&
v_+(n)\end{array}\right), \ee where the last equality follows from
Eq.~(\ref{as-2}). Therefore, in accordance with Eqs.~(\ref{recon})
and~(\ref{RH-3}), the solution $u_n(t)$ of the AL equation can be
retrieved from the solution of the RH problem as
\be\label{recon-2}
u_n(t)=-\lim_{z\to\infty}\frac{(z\Psi_+)_{12}}{\Psi_{+22}}=-\frac{\Psi_{+12}^{(1)}}
{\Psi_{+22}^{(0)}}=-\frac{\Gamma_{12}^{(1)}}{v_+(n)}. \ee

The matrix $\Gamma$ is mainly determined by the vector
$|y_j\rangle$. Now we derive a coordinate dependence of the
vector. It follows from the spectral problem that
\begin{eqnarray*}
& &\Psi_+(n+1,z_j)|y_j,n+1,t\rangle=0 \\
&=&\left(E(z_j)+Q_n\right)\Psi_+(n,z_j)E^{-1}(z_j)
|y_j,n+1,t\rangle.
\end{eqnarray*}
Hence, we can pose $E^{-1}(z_j)|y_j,n+1,t\rangle=|y_j,n,t\rangle$,
or \be\label{n-dep} |y_j,n,t\rangle=E^n(z_j)|y_j,t\rangle, \ee
where the vector $|y_j,t\rangle$ does not depend on $n$.
Similarly, it follows from Eqs.~(\ref{AL-temp}) and (\ref{vec-1})
that $|y_j,n,t\rangle_t=\Omega(z_j)|y_j,n,t\rangle$. Therefore,
the coordinate dependence of $|y_j,n,t\rangle$ is given as
\be\label{vec} |y_j,n,t\rangle=E^n(z_j)e^{\Omega(z_j)t}|p\rangle,
\qquad |p\rangle=\mathrm{const}. \ee

Finally, we find from the identity $\det\Psi_+(z_j,n,t)=0$ that
zeros $z_j$ do not depend on $n$ and $t$. Zeros $\pm z_j$,
$\pm\bar z_j$ and vectors $|y_j,n,t\rangle$ comprise the discrete
part of the RH problem data, while the functions $b_\pm(z)$
entering the matrix $G$~(\ref{RH-2}) are responsible for the
continuous data with the dependence on $t$ of the form
\be\label{G} G_t=[\Omega,G]. \ee

\section{Soliton solution}\label{sec:5}

In what follows we will not dwell on evident formulas for the
general case of $4N$ zeros. Instead we will give a detailed
account of the case of four zeros corresponding to a soliton (Fig.
1).
\begin{figure}
\includegraphics[scale=0.95]{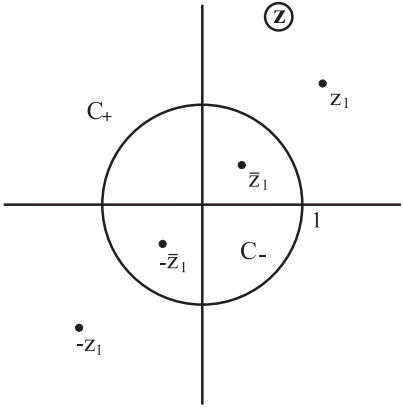}
\caption{Typical arrangement of zeros corresponding to a single
soliton. } \label{fig1}
\end{figure}
 Hence, after the regularization of the matrix RH problem with
zeros  $\pm z_1$ and $\pm\bar z_1$, we arrive at the regular RH
problem~(\ref{RH-4}). Solitons are associated with the discrete
part of the RH data, while the continuous data are now trivial
($G=\openone$). Hence, the solutions of the regular RH problem are
written in accordance with Eq.~(\ref{RH-reg}) as
\be\label{RH-reg-1} \psi_+(n)=\psi_-(n)=\left(\begin{array}{cc} 1 & 0\\
0& v_+(n)\end{array}\right). \ee

It is possible to express $v_+(n)$ (and $v_-(n)$) through
$\Gamma(n,z=0)$. Indeed, because now $\Psi_+=\Psi_-$, we find from
Eq.~(\ref{as-2}) $\Psi_+\longrightarrow\mathrm{diag}(v_-(n),1)$ as
$z\to 0$ which gives from $\Psi_+=\psi_+\Gamma$ and
Eqs.~(\ref{vec}) and~(\ref{B}):
\begin{eqnarray}\label{Gamma0}
\Gamma(n,0)&=&\mathrm{diag}(v_-(n),v_+^{-1}(n)), \\
B&=&\mathrm{diag}(\Gamma_{11}(n,0),\Gamma_{22}(n,0)^{-1}).
\nonumber\end{eqnarray} Thus, the reconstruction
formula~(\ref{recon-2}) for solitons is written more conveniently
as \be\label{recon-3}
u_n(t)=-\Gamma_{12}^{(1)}(n)\Gamma_{22}(n,0). \ee Hence, it is the
matrix $\Gamma(n,z)$ that determines the soliton solution. For
simplicity, we will hereafter denote the vector $|y_1,n,t\rangle$
as $|n\rangle$ with $|y_{-1},n,t\rangle=\sigma_3|n\rangle$.
Denoting $z_1=\exp[(1/2)(\mu+ik)]$ and
$(p_1/p_2)=\exp(a+i\varphi)$, where $p_1$ and $p_2$ are components
of the constant vector $|p\rangle$, we find from~(\ref{vec}) the
vector $|n\rangle$ explicitly: \be\label{vec-sol}
|n\rangle=e^{\frac{1}{2}(a+i\varphi)}\left(\begin{array}{l}
e^{\frac{1}{2}(x_n+i\varphi_n)}\\
e^{-\frac{1}{2}(x_n+i\varphi_n)}\end{array}\right). \ee Here
$x_n=\mu n-2t\sinh\mu\sin k+a$, $\varphi_n=kn+2t(\cosh\mu\cos
k-1)+\varphi$.

As regards the matrix $\Gamma$, it follows from
Eq.~(\ref{Gamma-1}) with $N=1$, $z_2=-z_1$ and $\bar z_2=-\bar
z_1$ that \begin{eqnarray}\label{Gamma-expl} & &\Gamma(n,z)
=\openone\\
&-&\frac{1}{z-\bar z_1}\left[|n\rangle(D^{-1})_{11})\langle
n|B+\sigma_3|n\rangle(D^{-1})_{21}\langle
n|B\right]\nonumber\\
&-&\frac{1}{z+\bar z_1}\left[|n\rangle(D^{-1})_{12}\langle
n|B\sigma_3+\sigma_3|n\rangle(D^{-1})_{22}\langle
n|B\sigma_3\right]. \nonumber\end{eqnarray} Calculating then
matrix elements $D_{kj}$~(\ref{D}) with
$\det\Gamma(n,0)=\exp(2\mu)$, we obtain from
Eq.~(\ref{Gamma-expl}) \be\label{Gamma-sol}
\Gamma(n,z)=\!\!\openone-\frac{\sinh\mu}{2(z-\bar z_1)}{\tilde
F_-}(n)-\frac{\sinh\mu}{2(z+\bar z_1)}{\tilde F_+}(n), \ee
\[
\Gamma^{-1}(n,z)=\!\!\openone+\frac{\sinh\mu}{2(z-z_1)}F_-(n)
+\frac{\sinh\mu}{2(z+z_1)}F_+(n), \
\] where
\[
{\tilde F_-}(n)\!=\!\!\left(\begin{array}{cc}
\frac{\exp\left[\mu(n-\frac{1}{2}-x)+\frac{i}{2}k\right]}{\cosh\mu(n-1-x)}
& \frac{\exp\left[ik(n-x)+i\alpha-\mu\right]}{\cosh\mu(n-1-x)}\\
\frac{\exp\left[-ik(n-1-x)-i\alpha+\mu\right]}{\cosh\mu(n-x)} &
\frac{\exp\left[-\mu(n-\frac{1}{2}-x)+\frac{i}{2}k\right]}{\cosh\mu(n-x)}
\end{array}\right),
\]
\[
F_-(n)\!=\!\!\left(\begin{array}{cc}
\frac{\exp\left[\mu(n-\frac{1}{2}-x)+\frac{i}{2}k\right]}{\cosh\mu(n-x)}
& \frac{\exp\left[ik(n-x)+i\alpha-\mu\right]}{\cosh\mu(n-1-x)}\\
\frac{\exp\left[-ik(n-1-x)-i\alpha+\mu\right]}{\cosh\mu(n-x)} &
\frac{\exp\left[-\mu(n-\frac{1}{2}-x)+\frac{i}{2}k\right]}{\cosh\mu(n-1-x)}
\end{array}\right),
\]
\be\label{D-sol} {\tilde F_+}(n)=-\sigma_3{\tilde F_-}(n)\sigma_3,
\quad F_+(n)=-\sigma_3F_-(n)\sigma_3. \ee Here
\begin{eqnarray}\label{coord}
x(t)&=&2t\frac{\sinh\mu}{\mu}\sin k+x_0, \quad
x_0=-\frac{a}{\mu}-\frac{3}{2}, \\
\alpha(t)&=&2t(\cosh\mu\cos k+\frac{k}{\mu}\sinh\mu\sin
k-1)+\alpha_0, \nonumber\\
\alpha_0&=&\varphi-\frac{ak}{\mu}-k.\nonumber
\end{eqnarray}
As a result, we obtain from Eq.~(\ref{recon-3}) the AL soliton
solution \cite{AL}: \be\label{sol}
u_n(t)=\exp[ik(n-x)+i\alpha]\frac{\sinh\mu}{\cosh\mu(n-x)}. \ee
Here and in what follows we write for simplicity $\cosh[\mu(n-x)]$
as $\cosh\mu(n-x)$. The AL soliton depends on four constant
parameters $\mu$, $k$, $x_0$ and $\alpha_0$ which determine
soliton mass $2\mu$, its group velocity
$v_{gr}=2(\sinh\mu/\mu)\sin k$, soliton maximum position $x(t)$
and phase $\alpha(t)$.%, the phase velocity being given by
%$v_{ph}=2(1-\cosh\mu\cos k)k^{-1}$.

It should be noted for later use that in the presence of a
perturbation Eqs.~(\ref{coord}) are modified due to possible
perturbation-induced evolution of the soliton parameters:
\begin{eqnarray}\label{pert-coor}
x(t)&=&\frac{2}{\mu}\int^t\sinh\mu(t')\sin
k(t')\mathrm{d}t'+x_0(t),\\
\alpha(t)&=&2\int^t[\cosh\mu(t')\cos k(t')-1]\mathrm{d}t'
\nonumber\\
&+&2\frac{k}{\mu}\int^t\sinh\mu(t')\sin
k(t')\mathrm{d}t'+\alpha_0(t).\nonumber
\end{eqnarray}

\section{Perturbation-induced evolution of the RH data: exact
results}\label{sec:6}

Having formulated the basic ingredients of the RH approach to the
AL system, we now proceed to the consideration of the perturbed AL
equation \be\label{AL-pert}
iu_{nt}+u_{n+1}+u_{n-1}-2u_n+|u_n|^2(u_{n+1}+u_{n-1})=\epsilon
r_n. \ee The small parameter $\epsilon$ characterizes the
perturbation amplitude and $r_n$ describes the functional form of
the perturbation. To find corrections to the soliton caused by a
perturbation, we first derive the corresponding evolution of the
RH data. In order to distinguish between the 'integrable'  and
'perturbative' contributions to the evolution equations, we will
assign the variational derivative $\delta/\delta t$ to the latter.
For example, we write $i\delta u_n/\delta t=\epsilon r_n$, as
follows from Eq.~(\ref{AL-pert}), or, in matrix form,
\be\label{var} i\frac{\delta Q_n}{\delta t}=\epsilon\hat R_n,
\qquad \hat R_n=\left(\begin{array}{cc} 0 & r_n\\ r^*_n
&0\end{array}\right). \ee

\subsection{Continuous data}

Consider the spectral problem~(\ref{AL-spec}). The perturbation
causes a variation $\delta Q_n$ of the potential which in turn
leads to a variation of the Jost solutions. It follows from
Eq.~(\ref{AL-spec}) that these variations are written in the form
\begin{eqnarray}\label{J-var}
& &E^{-n}J_-^{-1}(n)\delta J_-(n)E^n \\
&=&\sum_{l=-\infty}^{n-1}E^{-(l+1)}J_-^{-1}(l+1)\delta
Q_lJ_-(l)E^l, \nonumber\\
& &E^{-n}J_+^{-1}(n)\delta J_+(n)E^n
\nonumber\\
&=&-\sum_{l=n}^\infty E^{-(l+1)}J_+^{-1}(l+1)\delta
Q_lJ_+(l)E^l,\nonumber
\end{eqnarray}
where $\delta Q_l=(\delta Q_l/\delta t)\delta t$ and we have used
$\delta J_\pm(n)\to 0$ as $n\to\pm\infty$. Hence, due to the
definition~(\ref{scatt}), we obtain from Eq.~(\ref{J-var}) a
variation of the scattering matrix: \be\label{S-var} \frac{\delta
S}{\delta t}=-i\epsilon S_+\Upsilon_+(z)S_-^{-1}=-i\epsilon
T_+^{-1}\Upsilon_-(z)T_-. \ee Here the matrices $S_\pm$ and
$T_\pm$ are defined in Eqs.~(\ref{s+}) and~(\ref{det-}) and we
introduce the matrix function \begin{eqnarray}
\Upsilon_\pm(N_a,N_b)&=&\sum_{l=N_a}^{N_b}E^{-(l+1)}\Psi_\pm^{-1}(l+1)\hat
R_l\Psi_\pm(l)E^l, \nonumber\\
\Upsilon_\pm(z)&=&\Upsilon_\pm(-\infty,\infty).
\label{Up+}\end{eqnarray} Note that they are the analytical
solutions $\Psi_\pm$ that enter naturally the matrices
$\Upsilon_\pm$.

Now we derive a variation of $\Psi_+$. We have from Eq.~(\ref{s+})
that $\delta\Psi_+=\delta J_+E^nS_+E^{-n}+J_+E^n\delta S_+E^{-n}$.
The first term in r.h.s. is transformed  to
$i\epsilon\Psi_+(n)E^n\Upsilon_+(n,\infty)E^{-n}\delta t$ by means
of Eq.~(\ref{J-var}), while the second term, due to
Eq.~(\ref{S-var}) and a trick $\delta S_+=\delta SM_{11}$,
$M_{11}=\mathrm{diag}(1,0)$, is written as
$-i\epsilon\Psi_+(n)E^n\Upsilon_+(z)M_{11}E^{-n}\delta t$.
Therefore, the variation of $\Psi_+(n)$ takes the form
\be\label{evol-1} \frac{\delta\Psi_+(n,z)}{\delta
t}=-i\epsilon\Psi_+(n,z)E^n\Pi_+(n,z)E^{-n}, \ee where $\Pi_+$ is
the evolution functional \cite{My} defined here by \be\label{Pi+}
\Pi_+(n,z)=\left(\begin{array}{ll}\Upsilon_{+11}(-\infty,n-1) &
-\Upsilon_{+12}(n,\infty)\\ \Upsilon_{+21}(-\infty,n-1) &
-\Upsilon_{+22}(n,\infty)\end{array}\right). \ee Therefore, in the
case of perturbations the evolutionary equation for $\Psi_+$ gains
the additional term responsible for the perturbation:
\be\label{evol+}
\Psi_{+t}=V\Psi_+-\Psi_+\Omega-i\epsilon\Psi_+E^n\Pi_+E^{-n}. \ee
Similarly, \be\label{evol-2} \frac{\delta\Psi_-^{-1}}{\delta
t}=i\epsilon E^n\Pi_-E^{-n}\Psi_-^{-1}, \ee with \be\label{Pi-}
\Pi_-(n,z)=\left(\begin{array}{cc} \Upsilon_{-11}(-\infty,n-1) &
\Upsilon_{-12}(-\infty,n-1) \\ -\Upsilon_{-21}(n,\infty) &
-\Upsilon_{-22}(n,\infty)\end{array}\right) \ee and
\be\label{evol-}
\Psi_{-t}^{-1}=-\Psi_-^{-1}V+\Omega\Psi_-^{-1}+i\epsilon
E^n\Pi_-E^{-n}\Psi_-^{-1}. \ee Remarkably, the functions
$\Upsilon_\pm$ are interrelated by the matrix $G$ entering the RH
problem: \be\label{Up+-} \Upsilon_-=G\Upsilon_+G^{-1}. \ee

The evolution functionals $\Pi_\pm$ play the key role in the
analysis of a perturbation because they contain all needed
information about it \cite{My}. It is seen from the
definitions~(\ref{Up+}), (\ref{Pi+}) and~(\ref{Pi-}) that the
matrices $\Pi_\pm$ are meromorphic (and $E^n\Pi_\pm E^{-n}$ are
bounded) in $\mathcal{C}_\pm$  having simple poles at zeros of
$\det\Psi_+(z)$ and $\det\Psi_-^{-1}$, respectively. Further, the
evolution equation for $G$ follows easily from Eqs.~(\ref{RH-1}),
(\ref{evol+}) and~(\ref{evol-}) and takes the form
\be\label{evol-G} G_t=[\Omega,G]-i\epsilon(G\Pi_+-\Pi_-G), \ee or,
for $\tilde G=\exp(-\Omega t)G\exp(\Omega t)$, \be\label{evol-con}
\tilde G_t=-i\epsilon(\tilde Ge^{-\Omega t}\Pi_+e^{\Omega
t}-e^{-\Omega t}\Pi_-e^{\Omega t}\tilde G). \ee

\subsection{Discrete data}

In the point $z_1$ \be\label{vec-3} \Psi_+(n,z_1)|n\rangle=0 \ee
and near this point \be\label{Pi+sum}
\Pi_+(z)=\Pi_+^{(\mathrm{reg})}(z)+\frac{1}{z-z_1}\mathrm{Res}_{z=z_1}\Pi_+(z),
\ee where $\Pi_+^{(\mathrm{reg})}$ is the regular part of $\Pi_+$
in the point $z_1$ and $\mathrm{Res}_{z=z_1}$ stands for the
residue at $z=z_1$. It is shown in the Appendix A that the
evolution equation for the eigenvector takes the form
\be\label{vec-evol} \vert n\rangle_t=\Omega(z_1)\vert
n\rangle+i\epsilon
E^n(z_1)\Pi_+^{(\text{reg})}(z_1)E^{-n}(z_1)\vert n\rangle. \ee
Remember that the $n$-dependence of $\vert n\rangle$ is given by
Eq.~(\ref{n-dep}), $\vert n\rangle=E^n(z_1)\vert \tilde p\rangle$,
with the $n$-independent vector $\vert \tilde p \rangle$.
Therefore, the perturbation-induced evolution of the vector $\vert
p\rangle=\exp[-\Omega(z_1)t]\vert \tilde p\rangle$ is governed by
the equation \be\label{54} \vert p\rangle_t=i\epsilon
e^{-\Omega(z_1)t}\Pi_+^{(\text{reg})}(z_1)e^{\Omega(z_1)t}\vert
p\rangle. \ee In the absence of perturbation, the vector $\vert
p\rangle$ in Eq.~(\ref{54}) coincides with that in
Eq.~(\ref{vec}).

Evolution of zero $z_1$ is derived by taking the total time
derivative of $\det\Psi_+(z_1)=0$. We obtain \[
z_{1t}=-\left[\frac{(\partial/\partial
t)\det\Psi_+(z)}{(\partial/\partial z)\det\Psi_+(z)}\right]_{z_1}.
\]
Because $(\partial/\partial
z)\det\Psi_+=-i\epsilon\mathrm{tr}\Pi_+\det\Psi_+$,
$\det\Psi_+=v_+(n)a_+(z)$ [see Eq.~(\ref{det+})] and
$a_+(z)=(z^2-z_1^2)(z^2-\bar z_1^2)^{-1}$, the latter formula
 following from $a_+(z)=a_+(-z)$, $\lim a(z)\to 1$ as
$z\to\infty$ and $a_+(\pm z_1)=0$, we obtain a simple equation
\be\label{evol-z}
z_{1t}=i\epsilon\mathrm{Res}_{z=z_1}\mathrm{tr}\Pi_+(n,z). \ee

It is important that l.h.s. of Eqs.~(\ref{54}) and~(\ref{evol-z})
do not depend on $n$. Therefore, we can consider these equations
for $n\to+\infty$ where
\[
\Pi_+(z)=\left(\begin{array}{cc}\Upsilon_{+11}(z) & 0\\
\Upsilon_{+21}(z) &0\end{array}\right).
\]
As a result, the evolution equations for the discrete RH data are
finally written as
\begin{eqnarray}
z_{1t}\!\!&=&\!\!i\epsilon\mathrm{Res}_{z=z_1}\Upsilon_{+11}(z), \label{56}\\
\vert p\rangle_t\!\!&=\!\!&i\epsilon
e^{-\Omega(z_1)t}\left(\begin{array}{cc}
\Upsilon_{+11}^{(\text{reg})}(z_1) &
0\\\Upsilon_{+21}^{(\text{reg})}(z_1)
&0\end{array}\right)e^{\Omega(z_1)t}\vert p\rangle.\label{57}
\end{eqnarray}

It should be noted that Eqs.~(\ref{evol-con}), (\ref{56})
and~(\ref{57}) are exact. However, they cannot be directly applied
because the matrices $\Pi_\pm$ and $\Upsilon_\pm$ depend on
unknown solutions $\Psi_\pm$ of the spectral problem with the
perturbed potential. In the following sections we will describe
for sufficiently small $\epsilon$ the iterative RH problem-based
procedure to consecutively account for two main approximations:
the leading-order adiabatic approximation and the next-order (the
first-order) one.

\section{Adiabatic approximation}\label{sec:7}

Within the adiabatic approximation, we ignore radiation effects
and assume that the soliton adjusts its hyperbolic secant shape to
perturbation at the cost of slow evolution of the parameters.
Evolution equations for the soliton parameters in the adiabatic
approximation have the form
\begin{eqnarray}
& &\mu_t=\epsilon\sinh\mu
\label{58}\\
&\times&\sum_{n=-\infty}^\infty\frac{\text{Im}(R_n)\cosh\mu(n-x)}
{\cosh\mu(n+1-x)
\cosh\mu(n-1-x)}, \nonumber\\
& & k_t=-\epsilon\sinh\mu \label{59}\\
&\times&\sum_{n=-\infty}^\infty\frac{\text{Re}(R_n)\sinh\mu(n-x)}
{\cosh\mu(n+1-x)
\cosh\mu(n-1-x)},\nonumber\\
& &x_t=\frac{2}{\mu}\sinh\mu\sin k +\frac{\epsilon}{\mu}\sinh\mu \label{60}\\
&\times&\sum_{n=-\infty}^\infty
\frac{(n-x)\text{Im}(R_n)\cosh\mu(n-x)}{
\cosh\mu(n+1-x) \cosh\mu(n-1-x)},\nonumber\\
& &\alpha_t=2(\cosh\mu\cos k+\frac{k}{\mu}\sinh\mu\sin
k-1)\label{61}
\\
&+&\epsilon\sum_{n=-\infty}^\infty\Bigl\{\bigl[(n-x)\sinh\mu\sinh\mu(n-x)\nonumber
\\
&-&\cosh\mu\cosh\mu(n-x) \bigl]\text{Re}(R_n)\nonumber\\
&+&\frac{k}{\mu}(n-x)\sinh\mu\cosh\mu(n-x)\text{Im}(R_n)\Bigl\}
\nonumber \\
&\times&\text{sech}\mu(n+1-x)
\text{sech}\mu(n-1-x).\nonumber\end{eqnarray} Here
$R_n=r_n\exp[-ik(n-x)-i\alpha]$ and $r_n$ is constructed by means
of the AL soliton solution~(\ref{sol}) .
Eqs.~(\ref{58})-(\ref{60}) have been obtained for the first time
in Ref.~\cite{V-G}. The derivation of Eqs.~(\ref{58})-(\ref{61})
within the RH problem approach is given in the Appendix B.

\section{Radiation effects}
\label{sec:8}

The continuous part of the RH data describes a distortion of the
soliton shape and emission of small-amplitude dispersive waves by
soliton. To account for the continuous data, we should abandon the
condition $G=\openone$ and admit a $z$-dependence of the regular
RH problem solutions $\psi_\pm$. In other words, we pose
\be\label{71} G=\openone+\epsilon g(z), \qquad
\psi_+(n,z)=\psi_+^0(n)(\openone+\epsilon\phi(n,z)), \ee where
$\psi_+^0$ stands for the solution~(\ref{RH-reg-1}) of the regular
RH problem~(\ref{RH-4}) in the adiabatic approximation, whereas
the off-diagonal matrices $g(z)$ and $\phi(z)$ describe
first-order corrections. Therefore, the reconstruction
formula~(\ref{recon-2}) takes now the form
\begin{eqnarray} \label{72}
u_n&=&-\lim_{z\to\infty}\frac{\left[z\psi_+^0(\openone+\epsilon\phi)
\Gamma\right]_{12}}{\left[\psi_+^0(\openone+\epsilon\phi)\Gamma\right]_{22}}
\\
&=&-\Gamma_{12}^{(1)}(n)\Gamma_{22}(n,0)-\epsilon\phi_{12}^{(1)}(n)\Gamma_{22}(n,0).
\nonumber\end{eqnarray} The first term in the r.h.s. of
Eq.~(\ref{72}) represents the familiar soliton solution in the
adiabatic approximation and the second one is responsible for
radiation (soliton shape distortion). For the derivation of
Eq.~(\ref{72}) we employ the fact that the off-diagonal matrix
$\phi$ satisfies the asymptotic condition $\phi\to
z^{-1}\phi^{(1)}+O(z^{-2})$ with
\[
\phi^{(1)}=\left(\begin{array}{cc}0 & \phi_{12}^{(1)} \cr
\phi_{21}^{(1)} & 0\end{array}\right).
\]

Evaluation of $\phi_{12}^{(1)}$ and hence of radiation corrections
to soliton solution reduces to solving the regular RH
problem~(\ref{RH-4}) with $G$ as in Eq.~(\ref{71}). Indeed, we
have $\psi_-^{-1}\psi_+=\openone+\epsilon\Gamma
E^ng(z)E^{-n}\Gamma^{-1}$ and the jump of the piece-wise
holomorphic function
$\psi(z)=\left\{\psi_+(z),z\in\mathcal{C}_+;\psi_-(z),z\in\mathcal{C}_-\right\}$
across the contour $\vert z\vert\!=\!1$ is written as
\be\label{73} \psi_+-\psi_-=\epsilon\psi_+^0\Gamma
E^ngE^{-n}\Gamma^{-1}. \ee Here we omit terms with higher order of
$\epsilon$ and invoke the equality $\psi_-^0=\psi_+^0$ [see
Eq.~(\ref{RH-reg-1})] valid in the adiabatic approximation. The
Plemelj formula gives for $z\in\mathcal{C}_+$:
\[
\psi_+(z)=\psi_+^0\left[\openone+\frac{\epsilon}{2\pi
i}\oint_{|z|=1}\frac{\textrm{d}z'}{z'-z}(\Gamma
E^ngE^{-n}\Gamma^{-1})(z')\right] .
\]
Inserting here $\psi_+$ from Eq.~(\ref{71}) and performing the
asymptotic expansion at $z\to\infty$, we obtain the expansion
coefficient \be\label{74} \phi^{(1)}(n)=-\frac{1}{2\pi
i}\oint_{|z|=1}\textrm{d}z(\Gamma E^ngE^{-n}\Gamma^{-1})(z) \ee
determining the radiation correction~(\ref{72}). Therefore, our
next step is concerned with finding the matrix $g$.

To this end, we turn to the evolution equation~(\ref{evol-con})
for the matrix $\tilde G$ which is evidently related to $g$:
\be\label{75} \tilde G=\openone+\epsilon\tilde g, \qquad \tilde
g=e^{-\Omega t}ge^{\Omega t}. \ee Substituting this equation into
 Eq.~(\ref{evol-con}), we obtain in the first order of
$\epsilon$ \be\label{76} i\tilde g_t=e^{-\Omega
t}(\Pi_+-\Pi_-)e^{\Omega t}. \ee Because $\tilde g$ does not
depend on $n$, we can put $n\to\infty$ in Eq.~(\ref{76}) which
gives
\[
\Pi_+(n\to\infty)-\Pi_-(n\to\infty)=\left(\begin{array}{cc}
\Upsilon_{+11}-\Upsilon_{-11} & -\Upsilon_{-12} \cr \Upsilon_{+21}
& 0\end{array}\right).
\]
Moreover, it follows from Eqs.~(\ref{Up+}) and (\ref{71}) that
$\Upsilon_-=\Upsilon_+$ in the first order of $\epsilon$. As a
result,
\[
i\tilde g_t=e^{-\Omega t}\left(\begin{array}{cc} 0 &
-\Upsilon_{+12} \cr \Upsilon_{+21} & 0\end{array}\right)e^{\Omega
t}
\]
and the equation for $\tilde g_{12}$ takes the form \be\label{78}
\tilde g_{12t}=i\exp\left[-i(z-z^{-1})^2t\right]\Upsilon_{+12}.
\ee It is important to stress that because $\Upsilon_{+12}$
corresponds to the first order correction, we can replace in the
definition~(\ref{Up+}) of $\Upsilon_+$ unknown solution $\psi_+$
of the regular RH problem~(\ref{73}) by the known one $\psi_+^0$.
Integrating then Eq.~(\ref{78}), we can find the matrix
$g$~(\ref{75}).

The further stage is to consider the integrand in Eq.~(\ref{74}).
It can be shown from
%\begin{eqnarray*}
$(\Gamma E^ngE^{-n}\Gamma^{-1})_{12}
=z^{-2n}\Gamma_{12}(\Gamma^{-1})_{12}g_{21}+z^{2n}
\Gamma_{11}(\Gamma^{-1})_{22}g_{12}$
%\end{eqnarray*}
and explicit expressions~(\ref{D-sol}) for $\Gamma$ that the term
with $g_{21}$ is multiplied by $\mathrm{sech}^2\mu(n-x-1)$ and
hence vanishes at $n\to\pm\infty$. As a result, we are left with
\be\label{79} I_{12}\equiv(\Gamma
E^ngE^{-n}\Gamma^{-1})_{12}=\left\{
\begin{array}{ll} \frac{z^2-z_1^2}{z^2-\bar
z_1^2}z^{2n}g_{12}, & n\to+\infty \\
\frac{z^2-\bar z_1^2}{z^2-z_1^2}z^{2n}g_{12}, & n\to-\infty
\end{array} \right. \ee

Let us summarize the main steps in calculating the radiation
correction for a given perturbation $r_n$. First, we should
explicitly find the function $\Upsilon_{+12}(z)$ from the
definition~(\ref{Up+}) with $\Psi_+=\psi_+^0\Gamma$, $\psi_+^0$
and $\Gamma$ being given in Eqs.~(\ref{RH-reg-1}) and
(\ref{D-sol}), respectively. Then we integrate Eq.~(\ref{78}) and
obtain the matrix $g$ given in Eqs.~(\ref{75}) and (\ref{71}). For
the known function $g_{12}(z)$ we obtain the integrand~(\ref{79}).
Finally, after calculating the integral~(\ref{74}) we arrive at
the needed result.

In the next section we illustrate the proposed formalism on an
example of calculating the radiation corrections to the AL soliton
in the case of some model perturbations.

\section{Examples}\label{sec:9}

Here we apply our formalism to describe the perturbed  AL soliton
dynamics for the typical representatives of dissipative and
conservative  perturbations - linear damping $r_n=-iu_n$ and
quintic perturbation $r_n=\vert u_n\vert^4u_n$. The interplay
between the dissipative and conservative perturbations for the AL
model is considered in the adiabatic approximation by Abdullaev
\textit {et al.} \cite{Abd} and numerically by Soto-Crespo \textit
{et al.} \cite{S-C}.

\subsection{Linear damping}

In this case $\mathrm{Re}R_n=0$,
$\mathrm{Im}R_n=-\sinh\mu\,\mathrm{sech}\mu(n-x)$ and we have in
the adiabatic approximation
\[
k=\mathrm{const}, \quad \sinh\mu=\sinh(\mu_0)e^{-2\epsilon t},
\quad \mu_0=\mu(t=0),
\]
\[
x_t=v_{gr}-\frac{2\pi\epsilon}{\mu^2}\frac{\tanh\mu}{\sinh(\pi^2/\mu)}\sin
2\pi x, \quad \delta\alpha/\delta t=kx_t.
\]
In the process of obtaining the equation for $x_t$ we use the
Poisson summation formula~\cite{M-F} \be\label{poisson}
\sum_{n=-\infty}^\infty
f(n\mu)=\frac{1}{\mu}\int_{-\infty}^\infty\textrm{d}y\,
f(y)\left[1+2\sum_{s=1}^\infty\cos\frac{2\pi sy}{\mu}\right] \ee
and, following Ref.~\cite{Kiv-1}, we restrict ourselves to the
linear harmonic term ($s=1$) only. Higher harmonics contain the
factor $\exp(-\pi^2s/\mu)$ which for $\mu\approx 1$ is evidently
small. Hence, mass of the soliton decreases exponentially,  its
group velocity acquires a constant value ($=2\sin k$) after some
transient period (Fig. 2), while its phase is governed by the
evolution of the soliton position $x(t)$.
\begin{figure}
\includegraphics[scale=0.8]{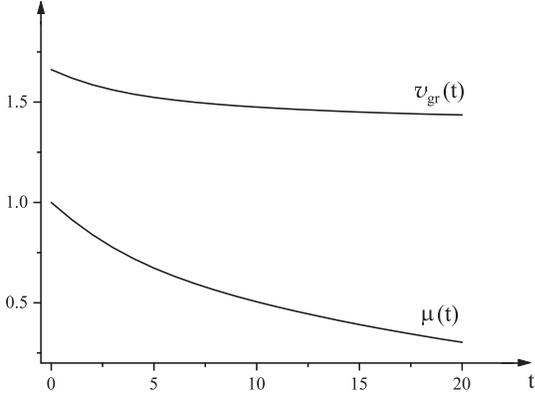} \caption{Evolution of the soliton
mass ($m=2\mu)$ and group velocity $v_{gr}$ in the case of linear
damping for $k=\pi/4$, $\epsilon=0.03$ and $\mu(t=0)=1$. }
\label{fig2}
\end{figure}

Now we embark on a calculation of radiation effects. Following the
prescriptions of Section \ref{sec:8} we find at first the matrix
function $\Upsilon_+$ written in accordance with Eq.~(\ref{Up+})
as \be\label{93} \Upsilon_+(z)=\sum_{n=-\infty}^\infty
E^{-(n+1)}(n+1)\Gamma^{-1}(n+1){\cal R}_n\Gamma(n)E^n. \ee Here
\begin{eqnarray*}
{\cal R}_n&=&(\psi^0_+)^{-1}(n+1)\hat
R\psi^0_+(n)\\
&=&\left(\begin{array}{cc} 0 & r_n\Gamma_{22}(n,0)^{-1}\\
r_n^*\Gamma_{22}(n+1,0) & 0
\end{array}\right)
\end{eqnarray*}
and
\[
\Gamma_{22}(n,0)=e^\mu\frac{\cosh\mu(n-x-1)}{\cosh\mu(n-x)}.
\]
Substituting $\Gamma$ and $\Gamma^{-1}$ (\ref{Gamma-sol}) into
Eq.~(\ref{93}), we arrive at
\begin{eqnarray*}
&&\Upsilon_{+12}=z^{-1}\frac{e^{-\mu+i\alpha-ikx}}{\cosh\mu-\cos(k-2\theta)}\\
&\times&\biggl[\bigl(1 -\cosh\mu\cos(k-2\theta)\bigr)S_1
+i\sinh\mu\sin(k-2\theta)S_2\biggr],
\end{eqnarray*}
where
\[
S_{\left\{1\atop 2\right\}}=\sum_{n=-\infty}^\infty
\frac{e^{ikn}z^{-2n}\left\{ \cosh\mu(n-x) \atop
\sinh\mu(n-x)\right\}}{\cosh\mu(n-x-1)\cosh\mu(n-x+1)}.
\]
Calculating these sums by means of the Poisson formula
(\ref{poisson}), we obtain a simple expression \be\label{102}
\Upsilon_{+12}=-\frac{\pi}{\mu z}\frac{\exp(-\mu+i\alpha-2i\theta
x)}{\cosh\mu\cosh(\pi(k-2\theta)/2\mu)}. \ee Here we pose
$z=\exp(i\theta)$ bearing in mind subsequent integration along the
contour $|z|=1$. What is more, because the radiation
correction~(\ref{72}) is multiplied by $\epsilon$, we restrict
ourselves to the leading term in each sum. Integrating then
Eq.~(\ref{78}) for $\tilde g_{12}$ with $\Upsilon_{+12}$ of the
form Eq.~(\ref{102}) and transforming the result to $g_{12}$ in
accordance with Eq.~(\ref{75}), we get
\[
g_{12}=-\frac{\pi}{\mu
z}\frac{e^{-\mu}}{\cosh\mu}\frac{\exp(i\alpha-2i\theta
x)}{\cosh(\pi(k-2\theta)/2\mu)}\frac{1-e^{-i\Lambda(\theta)t}}{\Lambda(\theta)}.
\]
Here $\Lambda(\theta)=(k-2\theta)v_{gr}+2(\cosh\mu\cos
k-\cos2\theta)$ and, within the first-order approximation, we can
take as $v_{ph}$ and $v_{gr}$ their initial values. Therefore, we
arrive at the integrand $I_{12}$~(\ref{79}) which determines the
integral~(\ref{74}). This integral can be calculated by residues.
The dominant contribution is provided by the third-order residue
in the point $z=\bar z_1$ (note that both
$\cosh(\pi(k-2\theta)/2\mu)$ and $\Lambda(\theta)$ have simple
zero in this point). The resulting expression is rather lengthy
and does not reproduced here. Instead we plot in Fig. 3 the
evolution of the perturbed AL soliton $u_n$~(\ref{72}) with
account for this expression.
\begin{figure} \vspace{-20mm}\centerline{}
\hspace{-15mm} \vspace{-5mm}
\includegraphics[scale=0.95]{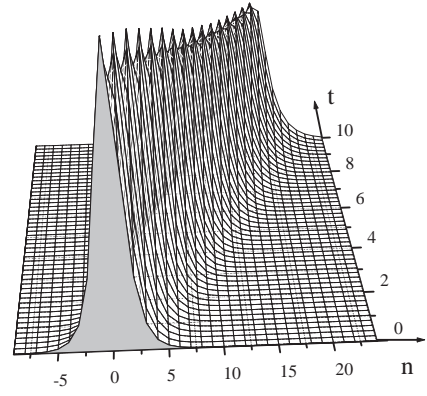}
\caption{Evolution of the perturbed AL soliton $u_n$~(\ref{72})
with account for the first-order correction. The soliton
parameters are the same as in Fig. 2. } \label{fig3}
\end{figure}
The shape of the soliton changes periodically due to the
discreteness of the system, with simultaneous decreasing of mass
in virtue of damping. More detailed conclusion about the
first-order contribution can be inferred from Fig. 4 where a
difference $|u_n|-|u_s|$ is pictured.
\begin{figure}
\includegraphics[scale=0.8]{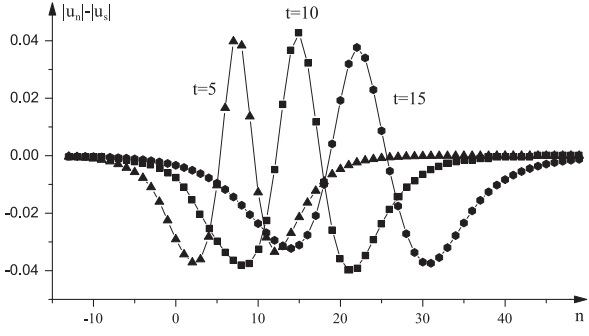}
 \caption{The difference $|u_n|-|u_s|$ for different time intervals
indicative of the shape distortion effect. Here $u_s$ is the
soliton in the adiabatic approximation and the soliton parameters
are the same as in Fig. 2. There are no long-lived nonvanishing
dispersive waves. } \label{fig4}
\end{figure}
Here $u_s$ is the AL soliton in the adiabatic approximation. The
shape distortion is mainly localized within the soliton 'envelope'
and in general is asymmetric. Such a behavior agrees with the
asymptotic representation of the first-order correction for
$n\to+\infty$:
\[
u_{n(\mathrm{rad})\!\!}=\!\!-\epsilon\phi^{(1)}_{12}\Gamma_{22}(n,0)
\approx-\frac{i\epsilon}{2H}\tanh\mu \,e ^{-\mu+i\alpha +ik}(n-x)
\]
\be\label{101} \times\left[n-x-\frac{\sin
k}{\sinh\mu}\left(1-\frac{\sinh2\mu}{2\mu}\right)H^{-1}\right]e^{(-\mu+ik)n}.
\ee Here $H=\sinh\mu\cos k-i\cosh\mu\sin k +(i/\mu)\sinh\mu\sin
k$. It follows from this expression that there are no nonvanishing
linear waves at $n\to\infty$. The similar result takes place for
$n\to-\infty$.

\subsection{Quintic Perturbation}

It follows from the results of the  Section \ref{sec:7} that for
\[
\textrm{Re}(R_n)=\sinh^5\!\mu\,\,\textrm{sech}^5\mu(n-x),\quad
\textrm{Im}(R_n)=0
\]
we have in the adiabatic approximation
\begin{eqnarray}\label{92}
&&\mu=\textrm{const}, \qquad x_t=2\frac{\sinh\mu}{\mu}\sin
k,\qquad
k_t=\\
&&\frac{2\epsilon\pi}{3\mu}\frac{\sinh^4\mu}{\sinh\frac{\pi^2}{\mu}}
\left[\frac{2\pi^4}{\mu^4}+\frac{2\pi^2}{\mu^2}\left(1-\frac{3}{\sinh^2\mu}\right)
+\frac{1}{3}\right]\sin 2\pi x. \nonumber
\end{eqnarray}
We do not write here the evolution equation for the phase $\alpha$
because of its lengthy form, though it is found quite immediately.
Hence, soliton mass is preserved in the presence of the quintic
perturbation, while the parameter $k(t)$ (and hence the group
velocity)  oscillates with a very small amplitude near the initial
value. Linear stability analysis demonstrates that the fixed
points of the evolutions~(\ref{92}) $k_s=\pi m, \quad
m=0,\pm1,\pm2,\ldots$, and $x_s=l/2,\quad l=0,\pm1,\pm2,\ldots$
are stable for even (odd) $m$ and odd (even) $l$.

Radiative corrections for the quintic perturbation are much the
same as for damping. Indeed, the function $\Upsilon_{+12}$ takes
the form
\begin{eqnarray*}
&& \Upsilon_{+12}=-\frac{\pi}{2\mu}e^{-\mu+i\alpha-2i\theta
 x}\frac{\sinh\mu}{\cosh(\pi(k-2\theta)/2\mu)}\\
 &\times&\left[P_3(\theta)-2e^{i(k-2\theta)}+\beta(\theta)P_4(\theta)\right],
 \end{eqnarray*}
where
\begin{eqnarray*}
&P_3(\theta)=\!\!\frac{1}{3}\left(\cosh\mu+i\frac{k-2\theta}{\mu}\sinh\mu\right)^3+\\
&\!\!\left(1-\frac{4}{3}\sinh^2\mu\right)\left(\cosh\mu+i\frac{k-2\theta}{\mu}\sinh\mu\right)
+\frac{2}{3}\cosh\mu,
\end{eqnarray*}
\[
P_4(\theta)=\frac{1}{24}\left(\frac{k-2\theta}{\mu}\sinh\mu\right)^4-\frac{1}{4}
\left(2+\frac{1}{3}\sinh^2\mu\right)
\]
\[
\times\left(\frac{k-2\theta}{\mu}\sinh\mu\right)^2+
\frac{1}{2}(1+\cosh^2\mu)-\cosh\mu\cos(k-2\theta),
\]
\[
\beta(\theta)=\left[\sinh\mu(\cosh\mu-\cos(k-2\theta))\right]^{-1}.
\]
Therefore, the integrand $I_{12}$ is written as
\[
I_{12}=-\frac{\pi}{2\mu}e^{-\mu+i\alpha}\frac{\sinh\mu}{\cosh(\pi(k-2\theta)/2\mu)}
\frac{z^2-z_1^2}{z^2-\bar z_1^2}z^{2(n-x)-1}
\]
\[
\times\left[P_3(\theta)-2e^{i(k-2\theta)}+\beta(\theta)P_4(\theta)\right]
\frac{1-e^{-i\Lambda(\theta)t}}{\Lambda(\theta)},
\]
with the same $\Lambda(\theta)$, as before. The result of
calculation of the integral (\ref{74}) with the above integrand
has the same structure as in Eq.~(\ref{101}). What is more, the
function $P_3(\theta)-2e^{i(k-2\theta)}$ being zero for $z=\bar
z_1$, does not contribute to the $n^2$-order of the radiative
correction.

\section{Conclusion}

We have proposed a formalism suitable for analytical investigation
of dynamics of the AL soliton   subjected to a perturbation. This
formalism provides a possibility of calculating both evolution of
the soliton parameters and perturbation-induced radiation effects.
Remarkably, it is the RH problem-based approach that has been
proved to be efficient for treating continuous nonlinear
equations, both integrable and nearly integrable, that turns out
to be the natural basis to study discrete nonlinear systems. We
have demonstrated within this approach how to consistently advance
from an integrable to perturbed system, in so doing the only
ingredient that should be added to the formalism to account for a
perturbation is the evolution functional $\Pi_+$ (or $\Pi_-$)
introduced by Shchesnovich \cite{My}. A natural step to further
extend the applicability of analytical methods in the the theory
of discrete nonlinear systems is to consider vector AL-type
solitons \cite{Abl}. Work in this direction is now in progress.

\begin{acknowledgments}

The authors are very grateful to V.S.~Gerdjikov and
V.S.~Shchesnovich for helpful discussions. This work was partly
supported by the Belarussian Foundation for Fundamental Research,
Grant No. $\Phi$01-052, and by UK-EPSRC, Grant No. GR/S47335/01.
\end{acknowledgments}

\appendix
\appendix
\section{Perturbed evolution of eigenvector}

In this Appendix we derive Eq.~(\ref{vec-evol}). Taking the total
time derivative of Eq.~(\ref{vec-3}) gives with account of
Eqs.~(\ref{evol+}) and~(\ref{Pi+sum}):
\begin{eqnarray}\label{50}
&&
\Bigl\{V(n)\Psi_+(n)-\Psi_+(n)\Omega-i\epsilon\Psi_+(n)\biggl[E^n\Pi_+^{(\mathrm{reg})}
E^{-n}\nonumber \\
&+&(z-z_1)^{-1}\mathrm{Res}_{z=z_1}(E^n\Pi_+E^{-n})\biggr]\nonumber\\
&+&z_t\frac{\partial}{\partial z}\Psi_+(n)\Bigr\}_{z=z_1}\vert
n\rangle +\Psi_+(n,z_1)\vert n\rangle_t=0.
\end{eqnarray}
Let us introduce a holomorphic function
$\tilde\Pi=-i\epsilon(z-z_1)E^n\Pi_+E^{-n}$ which evidently gives
\be\label{Pi-tilde} \tilde\Pi(z_1)\vert
n\rangle=-i\epsilon\mathrm{Res}_{z=z_1}\left[E^n\Pi_+(z)E^{-n}\right]\vert
n\rangle. \ee On the other hand, representing $\Pi_+(z)$  from
Eq.~(\ref{evol+})  as
\[
-i\epsilon
E^n\Pi_+(z)E^{-n}=\Psi_+^{-1}\Psi_{+t}-\Psi_+^{-1}V\Psi_++\Omega,
\]
we obtain \be\label{50b} \tilde\Pi(z_1)\vert
n\rangle=\left[(z-z_1)\Psi_+^{-1}\Psi_{+t}\right]_{z_1}\vert
n\rangle=-z_{1t}\vert n\rangle. \ee Comparing
Eqs.~(\ref{Pi-tilde}) and~(\ref{50b}), we arrive at \be\label{51}
i\epsilon\mathrm{Res}_{z=z_1}\left[E^n\Pi_+(z)E^{-n}\right]\vert
n\rangle=z_{1t}\vert n\rangle. \ee Applying now $\Psi_+(z_1)$ to
both sides of Eq.~(\ref{51}), we obtain the important identity
\be\label{iden}
\Psi_+(n,z_1)\mathrm{Res}_{z=z_1}(E^n\Pi_+E^{-n})=0. \ee
Eqs.~(\ref{51}) and (\ref{iden}) permit to considerably simplify
Eq.~(\ref{50}). Indeed, the last term in square brackets in
Eq.~(\ref{50}) is rearranged by means of Eq.~(\ref{iden}) as
\begin{eqnarray*}
%&-&i\epsilon\left[\Psi_+(z)\frac{1}{z-z_1}\mathrm{Res}_{z=z_1}(E^n\Pi_+(z)E^{-n})
%\right]_{z_1}\vert n\rangle\\
&-&i\epsilon\left[\frac{\Psi_+(z)-\Psi_+(z_1)}{z-z_1}
\mathrm{Res}_{z=z_1}(E^n\Pi_+(z)E^{-n}) \right]_{z_1}\vert
n\rangle \\
&=&-i\epsilon\left[ \frac{\partial}{\partial
z}\Psi_+(z)\right]_{z_1}\mathrm{Res}_{z=z_1}(E^n\Pi_+(z)E^{-n})\vert
n\rangle \\
&=&-z_{1t}\left[ \frac{\partial}{\partial
z}\Psi_+(z)\right]_{z_1}\vert n\rangle
\end{eqnarray*}
and cancels the same term in Eq.~(\ref{50}). As a result, the
evolution equation for the vector $\vert n\rangle$ takes the form
\[ \vert n\rangle_t=\Omega(z_1)\vert n\rangle+i\epsilon
E^n(z_1)\Pi_+^{(\text{reg})}(z_1)E^{-n}(z_1)\vert n\rangle. \]
{}

\section{Adiabatic Approximation}

Here we obtain within the RH problem approach
Eqs.~(\ref{58})-(\ref{61}) which govern the adiabatic dynamics of
the AL soliton.

First of all we turn to Eq.~(\ref{56}). In accordance with
Eqs.~(\ref{Up+}), (\ref{RH-3}), (\ref{RH-reg-1})
and~(\ref{Gamma0}) we write
\begin{eqnarray*}
&&  \mathrm{Res}_{z=z_1}\Upsilon_{+11}(z)
=\mathrm{Res}_{z=z_1}\\
&\times&\!\!\left[\frac{1}{z}\!\sum_{n=-\infty}
^\infty\!\Gamma^{-1}(n+1,z)\psi_+^{-1}(n+1)\hat
R_n\psi_+(n)\Gamma(n,z)\right]_{11}
\end{eqnarray*}
\[
=\frac{\sinh\mu}{2z_1}\Biggl[\sum_{n=-\infty} ^\infty F_-(n+1)
\]
\[
\times\left(\begin{array}{cc}0 & \Gamma_{22}^{-1}(n,0)r_n \\
\Gamma_{22}(n+1,0)r_n^*
&0\end{array}\right)\Gamma(n,z_1)\Biggr]_{11}.
\]
With account of explicit expressions~(\ref{Gamma-sol})
and~(\ref{D-sol}) for $F_-$ and $\Gamma$ we obtain
\begin{eqnarray*}
& &z_{1t}=-\frac{i\epsilon}{4}z_1\sinh\mu \\
&\times&\sum_{n=-\infty} ^\infty\frac{R_ne^{\mu(n-x)}
-R_n^*e^{-\mu(n-x)}}{\cosh\mu(n+1-x)\cosh\mu(n-1-x)},
\end{eqnarray*}
where $R_n=r_n\exp[-ik(n-x)-i\alpha]$. Then from the definition
$z_1=\exp[(1/2)(\mu+ik)]$ we easily derive Eqs.~(\ref{58})
and~(\ref{59}).

In order to obtain Eqs.~(\ref{60}) and~(\ref{61}), we should at
first calculate $\Upsilon^{(\text{reg})}(z_1)$:
\begin{eqnarray}
&\!&
\Upsilon^{(\text{reg})}(z_1)=\!\!\left[\Upsilon(z)-(z-z_1)^{-1}\mathrm{Res}_{z=z_1}
\Upsilon(z)\right]_{z_1}\label{A.1}\\
&=&\!\!\!\sum_{n=-\infty} ^\infty\!
E^{-(n+1)}\left(\openone+\frac{\sinh\mu}{4z_1}F_+(n+1)\right)\mathcal{R}_n
\Gamma(n,z_1)E^n \nonumber\\
&+&\lim_{z\to z_1}\frac{\sinh\mu}{2(z-z_1)}\sum_{n=-\infty}
^\infty\biggl[E^{-(n+1)}(z)F_-(n+1)\mathcal{R}_n \nonumber\\
&\times&\Gamma(n,z)E^n(z)%\nonumber
-E^{-(n+1)}(z_1)F_-(n+1)\mathcal{R}_n \Gamma(n,z_1)\biggr].
\nonumber
\end{eqnarray}
Here $\mathcal{R}_n=\psi_+^{-1}(n+1)\hat R_n\psi_+(n)$. The second
term in the r.h.s. of Eq.~(\ref{A.1}) gives the ratio $0/0$ in the
limit $z\to z_1$. Using the l'Hospital rule, we ultimately arrive
at
\begin{eqnarray*}
& &\Upsilon^{(\text{reg})}(z_1)=\sum_{n=-\infty} ^\infty
E^{-(n+1)}(z_1)\biggl\{\bigl(\openone \\
&-&\frac{\sinh\mu}{2z_1}\sigma_3F_-(n+1)
+\frac{\sinh\mu}{4z_1}F_+(n+1)\bigl)\mathcal{R}_n
\Gamma(n,z_1)\\
&+&\frac{\sinh^2\mu}{4}F_-(n+1)\mathcal{R}_n\left(\frac{\tilde
F_-(n)}{(z_1-\bar z_1)^2}+\frac{\tilde F_+(n)}{(z_1+\bar
z_1)^2}\right)\\
&-&n\frac{\sinh\mu}{2z_1}\left[\sigma_3,F_-(n+1)\mathcal{R}_n\Gamma(n,z_1)\right]
\biggr\}E^n(z_1).
\end{eqnarray*}
Therefore,
\begin{widetext}
\begin{eqnarray}
&\Upsilon^{(\textrm{reg})}_{11}(z_1)\!=\frac{1}{8}\!\sum_{n=-\infty}
^\infty\!\biggl[\bigl(3R_ne^{\mu(n-x)}
-R_n^*e^{-\mu(n-x)}\bigr)\sinh\mu +2\bigl(R_n\cosh\mu
+R_n^*e^{-\mu}\bigr)e^{\mu(n-x)}\nonumber\\&-4R_n\textrm{sech}
\mu(n+1-x)\biggr]
\textrm{sech}\mu(n+1-x)\textrm{sech}\mu(n-1-x),\label{A.2}
\end{eqnarray}
\be\label{A.3}
\Upsilon^{(\textrm{reg})}_{21}(z_1)=\frac{1}{4}\!\sum_{n=-\infty}
^\infty\!
\frac{z_1^{2n+1}e^{-ik(n-x)-i\alpha}}{\cosh\mu(n+1-x)\cosh\mu(n-1-x)}
\biggl[\left(\cosh\mu-\frac{3}{2}\sinh\mu\right)R_n \ee
\[
+\bigl(e^{-\mu}+2e^{-\mu(n-1-x)}\cosh\mu(n-x)
+\frac{1}{2}e^{-2\mu(n-x)}\sinh\mu\bigl)R_n^*
-2n\sinh\mu\left(R_n+e^{-2\mu(n-x)}R_n^*\right) \biggr]\!\! .
\]
\end{widetext}
Then we obtain from Eq.~(\ref{57}) the following evolution
equations for the components of the vector $\vert p\rangle$:
\begin{eqnarray*}
p_{1t}&=&i\epsilon\Upsilon^{(\text{reg})}_{11}(z_1)p_1, \\
p_{2t}&=&i\epsilon\Upsilon^{(\text{reg})}_{21}(z_1)\exp\left[i\int^t(z_1^2+z_1^{-2}-2)\text{d}t
\right]p_1
\end{eqnarray*}
Because $(p_1/p_2)=\exp(a+i\varphi)$, we have from
Eq.~(\ref{A.2}):
\begin{eqnarray*}
& &
\frac{\text{d}}{\text{d}t}(a+i\varphi)\\
&=&\frac{i\epsilon}{4}\sum_{n=-\infty}^\infty
\biggl[\left(3R_ne^{\mu(n-x)}-R_n^*e^{-\mu(n-x)}\right)\sinh\mu \\
&+&2n\left( R_ne^{\mu(n-x)}-R_n^*e^{-\mu(n-x)}\right)\sinh\mu \\
&-&2R_n^*e^\mu\cosh\mu(n-x)
\biggr]\\
&\times&\text{sech}\mu(n+1-x)\text{sech}\mu(n-1-x)
\end{eqnarray*}
which results in \begin{eqnarray}& & a_t=-\epsilon\sinh\mu
\sum_{n=-\infty}^\infty
\left(n+\frac{3}{2}\right)\label{A.4}\\
&\times&\frac{\text{Im}(R_n)
\cosh\mu(n-x)}{\cosh\mu(n+1-x)\cosh\mu(n-1-x)},
\nonumber\end{eqnarray}
\begin{eqnarray*}
\varphi_t\!\!&=&\!\!\epsilon\!\!\sum_{n=-\infty}^\infty\bigl[n\,\text{sech}\mu\,\text{sech}\mu(n-x)
-\cosh\mu\cosh\mu(n-x) \nonumber \\
&+&\frac{1}{2}\sinh\mu\,\sinh\mu(n-x)\bigr]\text{Re}(R_n)
\nonumber\\
&\times& \text{sech}\mu(n+1-x)\,\text{sech}(n-1-x). \label{A.5}
\end{eqnarray*}
It follows from Eqs.~(\ref{coord}) and~(\ref{pert-coor}) that
\begin{eqnarray*}
x_t&=&\frac{2}{\mu}\sinh\mu\,\sin
k-\frac{1}{\mu}\left[\left(x+\frac{3}{2}\right)\mu_t+a_t\right],
\\
\alpha_t&=&2\left(\cosh\mu\,\cos k+\frac{k}{\mu}\sinh\mu\sin
k-1\right)\\
&+&\left(x+\frac{1}{2}\right)k_t-\left(x+\frac{3}{2}\right)\frac{k}{\mu}\mu_t
-\frac{k}{\mu}a_t+\varphi_{t}.
\end{eqnarray*}
Inserting here Eqs.~(\ref{58}), (\ref{59}), (\ref{A.3})
and~(\ref{A.4}), we finally obtain Eqs.~(\ref{60}) and~(\ref{61}).

%\newpage
%\begin{center}
%CAPTIONS TO FIGURES
%\end{center}

%FIG. 1. Typical arrangement of zeroes corresponding to a single
%soliton.

%FIG. 2. Evolution of the soliton mass ($m=2\mu)$ and group
%velocity $v_{gr}$ in the case of linear damping for $k=\pi/4$,
%$\epsilon=0.03$ and $\mu(t=0)=1$.

%FIG. 3. Evolution of the perturbed AL soliton $u_n$~(\ref{72})
%with account for the first-order correction. The soliton
%parameters are the same as in FIG. 2.

%FIG. 4. The difference $|u_n|-|u_s|$ for different time intervals
%indicative of the shape distortion effect. There are no
%nonvanishing waves at infinities.

\end{document}